\documentclass[twoside]{ae100prg}
\usepackage{graphicx}
\usepackage[breaklinks]{hyperref}
\usepackage{booktabs}

\begin{document}
\title{Quantum fields in gravity}


\author{Giovanni Acquaviva$^1$}

\address{$^1$ Department of Physics, University of Trento (Italy)}

\email{acquaviva@science.unitn.it}

\begin{abstract}
We give a brief description of some compelling connections between general relativity and thermodynamics through $i)$ the semi-classical tunnelling method(s) and $ii)$ the field-theoretical modelling of Unruh-DeWitt detectors.  In both approaches it is possible to interpret some quantities in a thermodynamical frame.
\end{abstract}

\section{Introduction}

The idea of treating the emission of radiation from black holes as a tunelling process across the horizon traces back to the first path-integral derivation by Hartle and Hawking \cite{haw}. As a matter of fact, the null-geodesic method introduced by Kraus, Parikh and Wilczek \cite{kra,par} and the Hamilton-Jacobi method proposed more recently by Padmanabhan and collaborators \cite{sri} can be considered as semi-classical versions of the original derivation.  On the other hand, the Unruh-DeWitt detector \cite{unruh,dewitt} constitutes a field-theoretical approach to the problem, providing a more exact answer to questions regarding the particle content of the field and its thermal features for different observers.

\section{The tunnelling method(s)}

The \emph{null-geodesic} and the \emph{Hamilton-Jacobi} methods named in the introductive section both rely on the calculation of the classical action $S$ of a particle along a trajectory crossing the horizon.  Since such a trajectory is classically forbidden, the action itself developes an imaginary contribution which, in the WKB approximation, allows to calculate the tunnelling probability rate

\begin{equation}
 \Gamma_{em} \simeq \exp \left( -2\ \Im S  \right)\ \ \ ,
\end{equation}
where $\Im$ stands for imaginary part.  The use of Kodama-Hayward theoretical results \cite{kod,hay}, which allow to express observables of interest in terms of invariant quantities, has been a main ingredient.

\medskip\noindent

In \cite{acqua} this methods has been analysed in detail and the following results has been proven:

\begin{itemize}
 \item a solid basis for the covariance of the method has been given;
 \item formal equivalence of the two aforementioned approaches holds at least in stationary cases;
 \item the method provides an invariant and consistent answer in a variety of situations (higher-dimensional solutions, Taub and Taub-NUT solutions, decay of unstable particles, emission from cosmological horizons and naked singularities).
\end{itemize}

The calculation can be summarized in the following steps regarding the Hamilton--Jacobi approach:

\begin{enumerate}
 \item assume that the tunnelling particle's action $S$ satisfies the relativistic Hamilton--Jacobi equation
 \begin{equation}
  g^{\mu\nu} \partial_{\mu} S \partial_{\nu} S + m^2 = 0
 \end{equation}
 \item reconstruct the whole action, starting from the symmetries of the problem; the integration is carried along an oriented, null, curve $\gamma$ with at least one point on the horizon
 \begin{equation}
  S = \int_{\gamma} dx^{\mu}\ \partial_{\mu} S
 \end{equation}
 \item perform a near-horizon approximation and regularize the divergence in the integral according to Feynman's prescription: the solution of the integral has in general a non-vanishing imaginary part.
 \end{enumerate}

\noindent The result can be given in the general form

\begin{equation}
 \Gamma_{em} = \Gamma_{abs}\ \exp\left( -\frac{2\pi\, \omega_H}{\kappa_H} \right)
\end{equation}
\noindent
where $\omega_H$ and $\kappa_H$ are respectively the invariant energy of the tunnelling particle and the invariant surface gravity in Hayward's theory.  Through comparison of the transition rate with the Boltzmann factor, we can identify an invariant temperature
\begin{equation}
 T_H = \frac{\kappa_H}{2\pi}
\end{equation}

\section{Unruh-DeWitt detectors} 

We consider a conformally flat 4-dimensional metric, a massless scalar field conformally coupled to the metric and a two-level quantum system coupled the the scalar field.  The idea is to calculate the probability for the absorption of a scalar quantum and the consequent excitation of the two-level system through the \emph{transition rate}
\begin{eqnarray}\label{resp}
 \frac{dF}{d\tau}&=&\frac{1}{2\pi^2} \int_0^{\infty}\ \cos\left(E\, s\right)\, \left( \frac{1}{\sigma^2(\tau,s)} + \frac{1}{s^2} \right)\, ds\ + \nonumber \\
   &&\ \ - \frac{1}{2\pi^2} \int_{\Delta\tau}^{\infty} \frac{\cos \left( E\, s \right)}{\sigma^2(\tau,s)}
 \label{eq:equation2}
\end{eqnarray}
where $E$ is the energy gap of the detector and $s$ is the duration of the detection (see \cite{udw} for details on the construction of eq.\ref{resp}).  The second integral is the \emph{finite-time contribution}, generally an oscillating tail exponentially damped.  The bulk of the information about the transition rate comes from the geodesic distance between the ``switching on'' and ``switching off'' events, evaluated along a fixed trajectory $x(\tau)$
\begin{equation}
 \sigma^2(\tau,s) = a(\tau) a(\tau-s)\, \left[ x(\tau) - x(\tau-s) \right]^2
\end{equation}
whose inverse is proportional to the positive frequency Wightman function.  The $a(t)$ is the conformal factor.
\medskip\noindent

Let's analyze two simple stationary cases: the \emph{Schwarzschild black hole} and the \emph{de Sitter cosmology}.  The detector will be placed on a Kodama trajectory, which means that it will sit at fixed distance from the horizon.  Both cases can be treated in the same way, because the function $\sigma^2$ can be written in general
\begin{equation}\label{sigma}
 \sigma^2(s) = - \frac{4 V}{\kappa^2} \sinh^2\left( \frac{\kappa}{2 \sqrt{V}}\, s \right)
\end{equation}
where $\kappa$ is the surface gravity and $\sqrt{V} = \sqrt{-g_{00}}$.  A Wightman function which, as in eq.\ref{sigma}, is stationary and periodic in imaginary time is called ``thermal'' because when Fourier-transformed gives a Planckian transition spectrum.  In our case, calculating both the stationary and the finite-time contributions, 
\begin{eqnarray}
 \frac{dF}{d\tau}&=&\frac{1}{2\pi} \frac{E}{\exp\left( \frac{2\pi \sqrt{V} E}{\kappa} \right) - 1}\ +\nonumber \\
  && +\, \frac{E}{2 \pi^2} \sum_{n=1}^{\infty}\, \frac{n\, e^{-n \kappa \Delta\tau / \sqrt{V}}}{n^2 + V E^2/\kappa^2} \left( \frac{\kappa}{\sqrt{V} E} \cos(E \Delta\tau) - \sin(E \Delta\tau) \right)
\end{eqnarray}

\section{Conclusions}

As regards the tunnelling method, it has been shown that the formalism gives an invariant answer and allows extensions to more general black hole horizons in various dimensions as well as cosmological horizons and naked singularities.  Moreover, the extension to dynamical space-times has been carried out: in this framework the radiation seems to originate near the \emph{local trapping horizon}, not the global event horizon.\\
The Unruh-DeWitt detector constitutes a more exact approach to the Unruh-Hawking effect, relying on a quantum field-theoretical calculation.  In stationary cases the response function of the detector is shown to be thermal with temperature given by the surface gravity, just as in the tunnelling approach.  The generalization to non-stationary situations gives rise to problems in the analytical resolution and in general, when the background is time-dependent, the thermal interpretation seems lost

\section*{References}
\bibliography{acquavivabib}

\end{document}